\documentclass[sn-standardnature]{sn-jnl}

\newcommand{\aaa}{0.7}
\usepackage{amsmath}
\usepackage[]{hyperref}
\usepackage{overpic}

\jyear{2021}%

\theoremstyle{thmstyleone}%
\theoremstyle{thmstyletwo}%

\theoremstyle{thmstylethree}%

\begin{document}
	
	\title[TLC prediction]{High-throughput discovery of chemical structure-polarity relationships combining automation and machine learning techniques}
	
	
	\author[1,2]{\fnm{Hao} \sur{Xu}}
	\equalcont{These authors contributed equally to this work.}

	\author[1]{\fnm{Jinglong} \sur{Lin}}
	\equalcont{These authors contributed equally to this work.}
	
	\author[3]{\fnm{Qianyi} \sur{Liu}}
	
	\author[4]{\fnm{Yuntian} \sur{Chen}}
	
	\author[1]{\fnm{Jianning} \sur{Zhang}}
	
	\author[5]{\fnm{Yang} \sur{Yang}}
	
	\author[6]{\fnm{Michael C.} \sur{Young}}
	
	\author[7]{\fnm{Yan} \sur{Xu}}
	
	\author*[8,9]{\fnm{Dongxiao} \sur{Zhang}}\email{zhangdx@sustech.edu.cn}
	
	\author*[1]{\fnm{Fanyang} \sur{Mo}}\email{fmo@pku.edu.cn}

	\affil*[1]{\orgdiv{School of Materials Science and Engineering}, \orgname{Peking University}, \orgaddress{\street{} \city{Beijing}, \postcode{100871}, \state{} \country{P. R. China}}}
	
	\affil[2]{\orgdiv{BIC-ESAT, ERE, and SKLTCS, College of Engineering}, \orgname{Peking University}, \orgaddress{\street{} \city{Beijing}, \postcode{100871}, \state{} \country{P. R. China}}}
	
	\affil[3]{\orgdiv{College of Chemistry and Molecular Engineering}, \orgname{Peking University}, \orgaddress{\street{} \city{Beijing}, \postcode{100871}, \state{} \country{P. R. China}}}
	
	\affil[4]{\orgdiv{The EIT Institute for Advanced Study (EIAS)}, \orgname{Eastern Institute of Technology(EIT)}, \orgaddress{\street{} \city{Ningbo}, \postcode{315200}, \state{Zhejiang} \country{P. R. China}}}
	
	\affil[5]{\orgdiv{Department of Chemistry and Biochemistry}, \orgname{University of California Santa Barbara}, \orgaddress{\street{} \city{Santa Barbara}, \postcode{93106}, \state{CA}, \country{U.S}}}
	
	\affil[6]{\orgdiv{Department of Chemistry \& Biochemistry, School of Green Chemistry \& Engineering}, \orgname{The University of Toledo}, \orgaddress{\street{2801 W. Bancroft St.} \city{Toledo}, \postcode{43606}, \state{Ohio}, \country{U.S}}}
	
	\affil[7]{\orgdiv{Chemistry Service Unit}, \orgname{WuXi AppTec Headquarters}, \orgaddress{\street{} \city{Shanghai}, \postcode{200131}, \state{}, \country{P. R. China}}}
	
	\affil[8]{\orgdiv{Department of Mathematics and Theories}, \orgname{Peng Cheng Laboratory}, \orgaddress{\street{} \city{Shenzhen}, \postcode{518000}, \state{} \country{P. R. China}}}
	
	\affil[9]{\orgdiv{School of Environmental Science and Engineering}, \orgname{Southern University of Science and Technology}, \orgaddress{\street{} \city{Shenzhen}, \postcode{518055}, \state{} \country{P. R. China}}}
	
	
	\abstract{As an essential attribute of organic compounds, polarity has a profound influence on many molecular properties such as solubility and phase transition temperature. Thin layer chromatography (TLC) represents a commonly used technique for polarity measurement. However, current TLC analysis presents several problems, including the need for a large number of attempts to obtain suitable conditions, as well as irreproducibility due to non-standardization. Herein, we describe an automated experiment system for TLC analysis. This system is designed to conduct TLC analysis automatically, facilitating high-throughput experimentation by collecting large experimental data under standardized conditions. Using these datasets, machine learning (ML) methods are employed to construct surrogate models correlating organic compounds’ structures and their polarity using retardation factor (\textit{R$_f$}). The trained ML models are able to predict the \textit{R$_f$} value curve of organic compounds with high accuracy. Furthermore, the constitutive relationship between the compound and its polarity can also be discovered through these modeling methods, and the underlying mechanism is rationalized through adsorption theories. The trained ML models not only reduce the need for empirical optimization currently required for TLC analysis, but also provide general guidelines for the selection of conditions, making TLC an easily accessible tool for the broad scientific community.}
	
	
	\maketitle
	
	\section*{Introduction}
	
\begin{figure*}[h]%
	\centering
	\includegraphics[scale=\aaa]{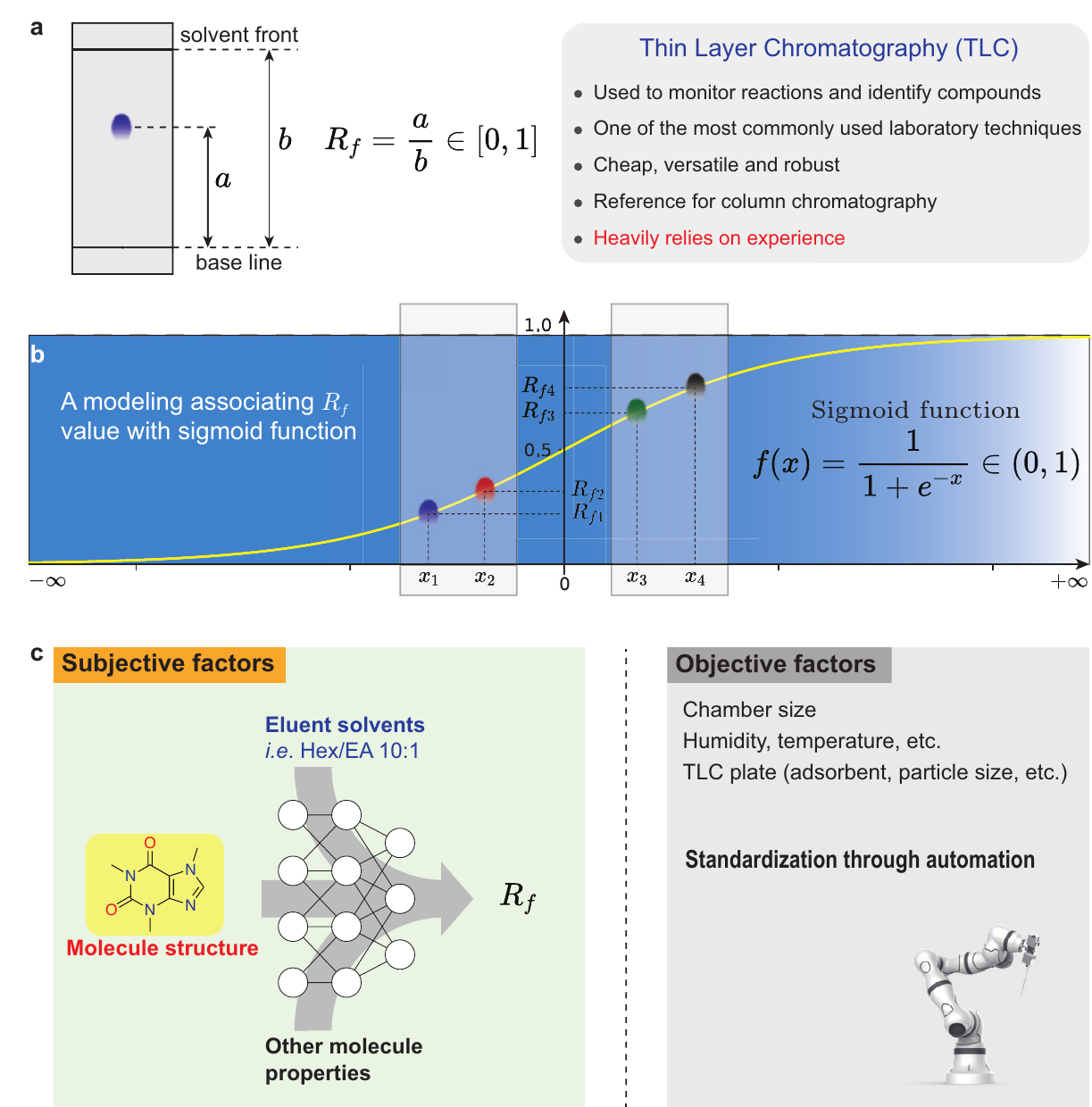}
	\caption{\textbf{Context of the work.} \textbf{a}, Thin-layer chromatography (TLC) is a chromatography technique used to separate non-volatile mixtures. Synthetic laboratories heavily use TLC techniques to monitor reactions and identify compounds daily. Choosing suitable TLC conditions is usually time-consuming for novices or for new compounds. The retardation factor (\textit{R$_f$}) is the fraction of an analyte in the mobile phase of a chromatographic system. It is defined as the ratio of the distance traveled by the center of a spot to the distance traveled by the solvent front. \textbf{b}, A sigmoid function is a mathematical function having a characteristic ``S"-shaped curve, and has domain of all real numbers with a return value in the range 0 to 1. Considering that the \textit{R$_f$} value also has the same value range, we deliberately associate it with sigmoid function. \textbf{c}, The subjective and objective factors to compound \textit{R$_f$} value measurement. The subjective factors include the compound’s structure and other physical properties, as well as elution solvents. The information can be mapped to a vector space via feature engineering, and then fed to ML algorithms. Other factors like chamber size, humidity, etc., can also affect the measurement. The influence of these objective factors should be eliminated as much as possible to avoid their impact on model training.}\label{fig1}
	\end{figure*}

	Thin layer chromatography (TLC) is a commonly used technique in modern chemistry and biology laboratories. As a key chromatography technique, the employment of a solid stationary phase and a liquid mobile phase allows for the separation of individual components of a complex mixture on the basis of their relative affinities for the two phases (Figure~\ref{fig1}a)\cite{sherma2003handbook}. TLC analysis is currently used routinely for reaction monitoring, product identification, and determination of chromatography conditions for subsequent purification. While highly experienced synthetic practitioners are able to use this tool, TLC techniques often present a significant hurdle for scientists in synthesis-adjacent fields. Furthermore, the identification of TLC conditions for new compound classes requires the judicious selection of several variables, most notably the mobile phases and their ratios, to achieve optimal separation. Traditionally, such goals are accomplished through trial-and-error in an extremely time and labor-intensive manner.
	
	In recent years, cutting-edge techniques in artificial intelligence (AI) have revolutionized the extrapolation of structure-property relationships in the chemical sciences\cite{muratov2020qsar}. In particular, machine learning (ML) algorithms are able to solve complex chemical problems with respect to prediction, surrogate model construction, and constitutive relationship discovery\cite{yang2020holistic, howarth2020dp4, reid2019holistic, kirkpatrick2021pushing}. In this vein, we postulated that a trained model might possibly work on predicting the polarity and specifically the \textit{R$_f$} value of organic compounds due to the strong structure-property relationship in the physical mechanism underlying TLC.
	
	Essentially, the task of prediction is to establish a certain mathematical model that maps between input and output. We noted that the sigmoid function, a common activation function in neural network algorithms, is tightly bound to the pair of horizontal asymptotes and highly useful in compressing or squashing outputs within a 0 to 1 range. The output of TLC, \textit{R$_f$}, is also a value between 0 and 1. As such, we envisioned that \textit{R$_f$} values could be reasonably fitted using a sigmoid function (Figure~\ref{fig1}b). In this case, all the factors that have an influence on \textit{R$_f$} value will be selected and subject to a matrix operation through certain input forms to determine an $x$, and the output, \textit{R$_f$} value, will be the sigmoid function value of the $x$. Additionally, in a reverse perspective, through such a transformative operation, the original range of 0 to 1 is extended to the entire real number domain. In other words, we can predict the \textit{R$_f$} value by regression in a larger range to improve the accuracy of the model (\textit{vide infra}).
	
	ML methods capture underlying patterns from a large amount of training data in order to make prediction. The availability of large, high-quality datasets is the prerequisite for ML methods. Although the \textit{R$_f$} value of newly-prepared organic compounds are available in the chemical literature, the lack of standardization often leads to inconsistent data, which impedes the development of ML models. Generating a sufficient amount of highly standardized data through conventional means is tedious and time-consuming. To address this challenge, we sought to exploit automated instrumentation to accelerate and standardize the measurement of \textit{R$_f$} values of organic compounds in TLC analysis. Furthermore, we use these data for ML model training to correlate compound structures and their polarity (Figure~\ref{fig1}c).
		
	\section*{Data acquisition}
		
	The automation of chemical experimentation has been an expanding field in the last decade\cite{ley2015organic, hase2019next, wilbraham2020digitizing}. These automated platforms enable standardization and enhances reproducibility while reducing experimental costs. Most importantly, automation allows for the generation of sufficient data for further statistical analysis. The research paradigm combining automation and ML techniques has already been successfully applied to chemical sciences in various scenarios ranging from reaction conditions optimization to mechanism interpretation\cite{ahneman2018predicting, granda2018controlling, steiner2019organic, burger2020mobile, newman2021univariate, shields2021bayesian}.
	
	We realized that the most effective strategy would be to develop an automated robotic platform to collect the necessary TLC data. However, TLC analysis requires several experimental steps, including dissolving the analyte, spotting the analyte, developing the TLC plate, and ultimately measuring and calculating the \textit{R$_f$} value for each analyte, thus demanding sophisticated automation.
	
	\begin{figure}[h]%
		\centering
		\includegraphics[scale=\aaa]{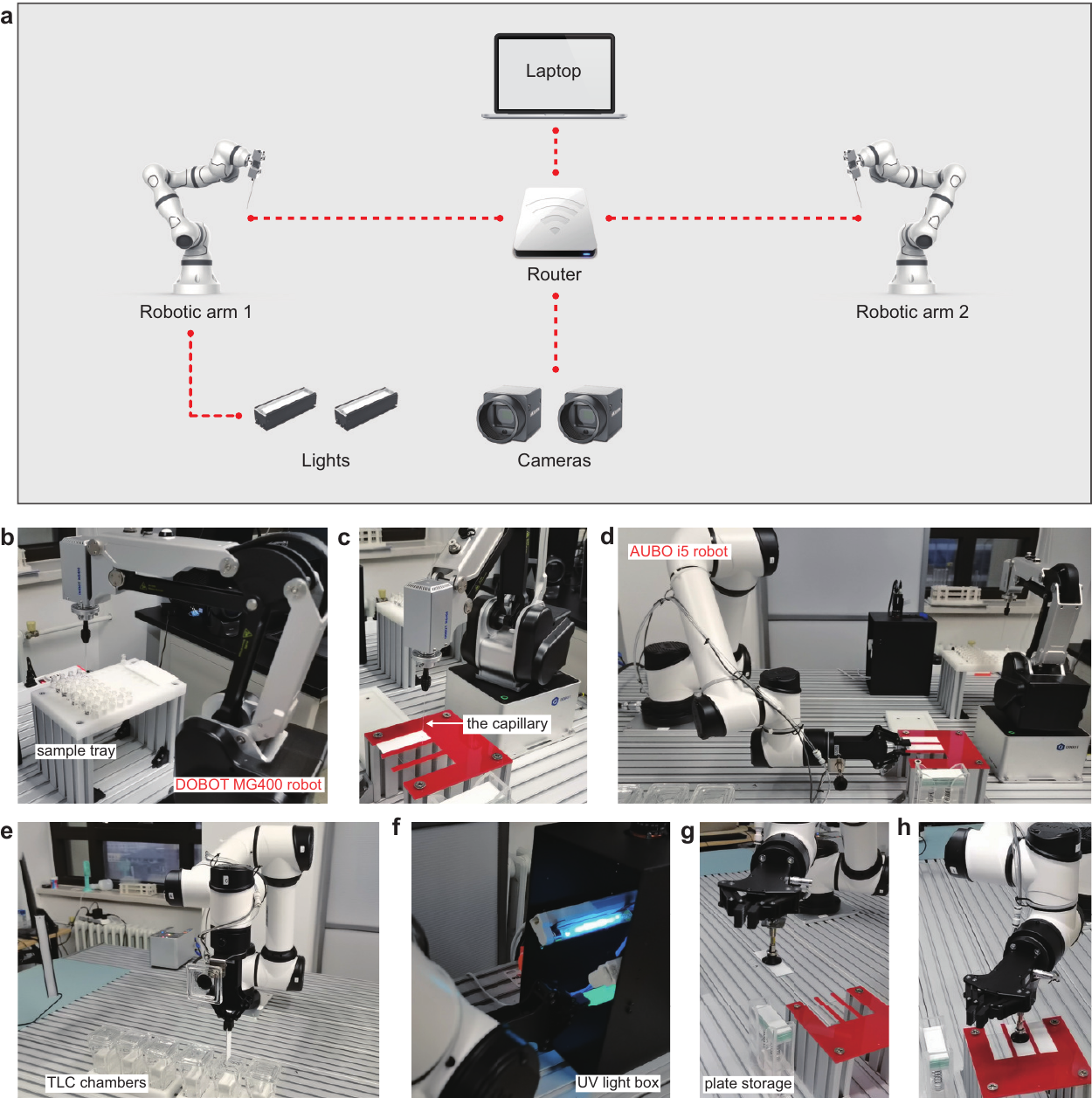}
		\caption{\textbf{Automated thin layer chromatography robots and experimental station.} \textbf{a}, Schematic of the automated robotic platform for TLC data collection. \textbf{b}, The DOBOT MG400 robot equipped with a capillary is drawing TLC samples on the sample tray. \textbf{c}, The DOBOT MG400 robot is spotting samples onto the TLC plates. \textbf{d}, The AUBO i5 robot is gripping a TLC plate prior to moving. \textbf{e}, The AUBO i5 robot is putting a TLC plate to TLC chamber (or retrieving a plate from the chamber). \textbf{f}, After developing, the AUBO i5 robot is taking the TLC plate to a UV light box to visualize and photograph. \textbf{g}, The AUBO i5 robot is retrieving a plate from storage. \textbf{h}, The AUBO i5 robot is placing a pristine TLC plate onto the stand.}\label{fig2}
	\end{figure}
	
	To address the challenges mentioned above, we built a set of custom desktop robot system for high-throughput collection of TLC data. Our robotic platform is shown in Figure~\ref{fig2}. Two collaborative robots are the core of this system, which are able to accomplish complex actions like human arms with high precision and safety. The system also comprises two cameras, two lights, a router, and a laptop (Figure~\ref{fig2}a). The smaller robot, DOBOT MG400, is responsible for drawing TLC samples on the sample tray (Figure ~\ref{fig2}b), and then spotting samples onto the TLC plates (Figure~\ref{fig2}c). Subsequently, the bigger robot, AUBO i5, is responsible for gripping the TLC plate, and then putting it into the chamber for development (Figure~\ref{fig2}d \& e). Upon completion, the TLC plate is transferred to a box for visualizing and photographing under UV light (Figure~\ref{fig2}f). Finally, the AUBO i5 will retrieve a pristine plate from the storage, and place it onto the stand for regenerating the system for the next usage (Figure~\ref{fig2}g \& h). In order to increase throughput and efficiency, six chambers are placed in the platform with different elution solvents in each. A Python program was built to control the robots, cameras and lights, thus managing the whole workflow. Through this robotic TLC experiment platform, we achieved high-throughput standardized polarity measurements.
	
	\section*{Polarity prediction with machine learning techniques}
	
\begin{figure}[H] 
		\centering
		\includegraphics[scale=\aaa]{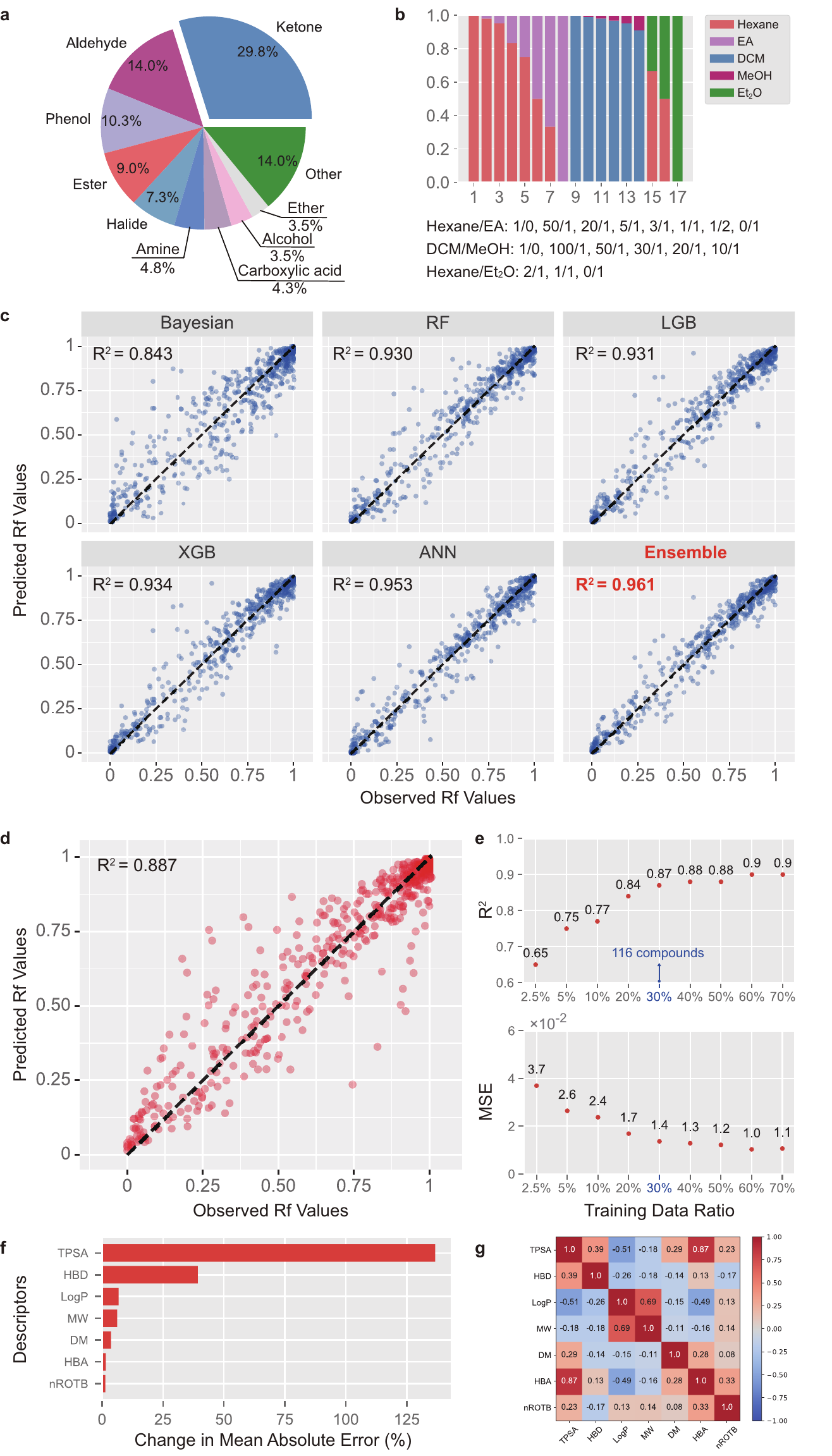}
		\caption{\textbf{Dataset description and polarity prediction with machine learning techniques. a}, The top 10 classes of the compounds in the dataset. \textbf{b}, The developing agent and their ratios utilized in this work. EA, ethyl acetate; DCM, dichloromethane; \textbf{c}, Observed versus predicted \textit{R$_f$} values for different ML methods. For all the models, an 80/10/10 random split of training, validation, and test data by TLC data was performed to measure the prediction ability of each model. Only test set data are shown in plots. The dashed line is $y = x$ line. \textbf{d}, Observed versus predicted \textit{R$_f$} values for the proposed ensemble method to predict out-of-sample compounds. The dataset is randomly split into 80/10/10 by compounds and only test set data are shown in plots. The dashed line is $y = x$ line. \textbf{e}, Test set performance of the ensemble prediction model with sparse data. The smaller training sets were selected randomly from the entire compounds and the test data are kept the same. A gradual erosion in predictive accuracy occurred from 70\% of the entire compounds down to 2.5\%.  \textbf{f}, The relative importance of the molecular descriptors utilized in this work. \textbf{g,} The heatmap of correlation coefficients between molecular descriptors.}\label{fig3} 
	\end{figure}

	After the data collection is completed by the automated TLC system, the \textit{R$_f$} values can be automatically calculated by an image analysis computer program, leading to highly standardized TLC runs. For UV-inactive compounds, other visualization methods can be adopted. In a typical task, the \textit{R$_f$} values of each compound under different elution solvent ratios are plotted, which helps chemists find outliers based on intuition and experience. Through this method, a high-quality dataset of compound \textit{R$_f$} values is established, containing 4944 standardized polarity measurements from 387 organic compounds (Figure~\ref{fig3}a) under three elution solvent systems including hexane/ethyl acetate, dichloromethane/methanol and hexane/diethyl ether with a total of 17 different solvent composition (Figure~\ref{fig3}b). These typical compounds were collected deliberately based on their types to get better representation.
	
	We next turned to selecting suitable molecular descriptors. In terms of mechanism, molecular polarity is a reflection of a compound’s polar bonds and their spatial distribution, thus essentially, its structure. In addition, TLC is based on the principle of separation through adsorption type. For this reason, polarity represented by \textit{R$_f$} value is also correlated with the properties that affect the molecular adsorption between stationary phase and mobile phase, for example, hydrogen bonding and topological polar surface area, etc. While many molecular fingerprints have been developed, we pay more attention to the ones that related to structure. Molecular Access System (MACCS) keys are one of the most commonly used structural keys that are employed to represent molecular structure, fragments, and substructural information\cite{durant2002reoptimization}. Each compound can be converted into MACCS keys with 167 dimensions. Meanwhile, molecular properties such as molecular weight (MW), topological polar surface area (TPSA), and many others, can also be conveniently extracted by using RDKit software\cite{landrum2013rdkit}. Moreover, another possible factor, dipole moment, was calculated using Mopac2016 (PM6-D3)\cite{stewart2007stewart}. For the elution solvents, vectorization encoding technique was used to express mobile phase information. The values in individual solvent column represent their ratio for a combination of an eluent. To avoid prohibitively time-consuming analysis and logging of computational data, we developed software to automate feature generation. The program requires only the input of SMILES string\cite{weininger1988smiles} of compounds in a Python script. The program then generates the data table that can be used for modeling. In total, 178-dimensional input metrics were extracted by the software to characterize each TLC set of conditions. One of the advantages of ML modeling is that one can define feature engineering relatively more freely without recourse to a specific hypothesis. This, to a certain extent, allows the machine to discover the connection, or even knowledge hidden in data without relying on human experience.
	
	With these data in hand, we evaluated the predictive accuracies of a series of ML methods including Bayesian regression, Random Forest (RF), LightGBM (LGB), XGBoost (XGB), and Artificial Neural Network (ANN). Here, the dataset is randomly divided into training, validation, and test set according to the ratio of 80\%, 10\%, and 10\% by TLC data. For ANN utilized in this work, the number of hidden layers is 4 with 128 neurons in each hidden layer, the number of input neurons is 178 and the number of output neurons is 1. Considering the physical constraint that the range of \textit{R$_f$} values is between 0 and 1, in the training process of the above-mentioned ML algorithms, the output will be mapped from the entire real number domain to [0,1] through the sigmoid function. It is discovered that deep learning-based algorithm like ANN and decision tree-based algorithms like RF, LGB and XGB all show satisfactory prediction ability with $R^2$ over 0.93 (Figure~\ref{fig3}c).  On the basis of these techniques, an ensemble method is proposed to form a better model and further improve the accuracy. We are delighted to find that a predictive model with the highest accuracy ($R^2 = 0.961$) can be obtained by a simple weighted average of these methods. This improvement may be attributed to the ablity of the ensemble algorithm to avoid overfitting and decrease the risk of obtaining a local minimum\cite{sagi2018ensemble}.
	
	Next, a more difficult and practical problem is considered that predicting the \textit{R$_f$} values of out-of-sample compounds under different solvent composition. In this task, the dataset is randomly split into a training dataset with 3922 \textit{R$_f$} values from 308 compounds (80\% of total compounds), a validation dataset with 484 \textit{R$_f$} values from 38 compounds (10\% of total compounds), and a test dataset with 473 \textit{R$_f$} values from 38 compounds (10\% of total compounds). It is worth mentioning that the test set is out-of-sample. The proposed ensemble model is trained and the prediction performance on the test set is examined. On average, the MSE of the predicted \textit{R$_f$} values of the out-of-sample 38 compounds was 0.117, with an $R^2$ value of 0.887 (Figure~\ref{fig3}d). The effective out-of-sample prediction of the proposed model suggests that the constitutive relationship between the input information and output polarity measurements were captured well by the prediction model, which means that it may be possible to predict the \textit{R$_f$} values of compounds under specified developing system without experiments.
	
	In order to explore the relationship between the predictive power of the model and input data more deeply, different numbers of compounds are utilized to train the model while the test set is fixed. It is surprising to discover that the prediction model is able to maintain satisfactory predictive power with a markedly smaller training data since the $R^2$ of the model trained from \textit{R$_f$} values of 116 compounds (30\% of total compounds) achieves 0.870, which is only a decrease of 0.03 compared with the model trained from 70\% of the total compounds (Figure~\ref{fig3}e). This indicates that there is a tight physical relationship between the structure and properties of the compound and the \textit{R$_f$} value that could be learned by the proposed ensemble model through only 116 compounds, so as to predict the polarity of countless kinds of organic compounds, which provide an excellent annotation of ``small data, big task"\cite{qi2020small}.
	
	After having obtained a satisfactory predictive model, we sought to explore the physical and chemical knowledge underlying the trained model. The molecular descriptors that characterize molecular properties are contained in the input information. As a consequence, the relative importance of utilized molecular descriptors are evaluated by the percent increase in the predictive model’s mean absolute error (MAE) when values of certain descriptors in the test set are randomly reassigned according to data distribution. It was found that, among these descriptors, TPSA shows significant importance over all others (Figure~\ref{fig3}f). The TPSA of a molecule is defined as the surface sum over all polar atoms, primarily oxygen and nitrogen, which is often used to evaluate the transport properties of drugs in cells, and it is proven to have an inseparable relationship with polarity in this work.  We rationalize that a larger TPSA leads to a stronger interaction between the adsorbate and the adsorbent (e. g. silica gel), leading to smaller $R_f$ values. As for HBD and HBA, they indicate the number of hydrogen bond donors and acceptors, respectively. Our results showed that HBD is more relevant to \textit{R$_f$} value over HBA. The explanation is that the solid phase used in this study is silica gel, which contains many bridge oxygens and hydroxyl groups on the surface. These surface oxygen atoms act as hydrogen bond accepters. As such, the analyte will have more binding action with silica gel when it contains hydrogen bond donors rather than hydrogen bond accepters. The correlation coefficients between molecular descriptors was also explored (Figure~\ref{fig3}g). It was found that TPSA and HBA have a very strong positive correlation.

	\section*{Conclusion}
	TLC experiments are performed thousands of times every day in synthetic laboratories around the world. Determination of suitable TLC conditions usually requires ample experience of chemists, and a large number of attempts would be inevitable. Here we have built a robotic experimentation platform and its software for TLC data collection, which generates a large amount of standardized compound’s structure-polarity data automatically. Using the dataset, we further developed ML models that can predict compound’s \textit{R$_f$} value with high accuracy. We expect that this TLC prediction approach will prove to be useful to the synthetic community in facilitating laboratory efficiency.
	
\backmatter

\bmhead{Supplementary information}

Supplementary files will be available along with the publication of  this article.

\bmhead{Acknowledgments}

We thank the Natural Science Foundation of China for funding support.

\section*{Declarations}
\bmhead{Funding} This work is supported by the Natural Science Foundation of China (Grant Nos. 22071004, 21933001, 22150013).
\bmhead{Competing interests}
F.M., H.X. and D.Z. are inventors on two patent applications (CN 202111638511.2 and 202122346010.9) submitted by Peking University that cover an organic chemistry laboratory automation system and a machine learning method for TLC conditions prediction, respectively.
\bmhead{Availability of data and materials}
The TLC dataset will be available along with the publication of  this article. A video associated with this project can be found via the following links:

English version: \url{https://www.bilibili.com/video/BV1am4y1o7yE/}

Chinese version: \url{https://www.bilibili.com/video/BV17R4y1j7jz/}

\bmhead{Code availability }
The code will be available along with the publication of  this article.
\bmhead{Authors' contributions}
 H.X. and F.M. built the robotics platform for high-throughput experimentation. J.L. performed TLC experiments. H.X. performed chemoinformatic and machine learning studies. Q.L. derived the theoretical formula for TLC mechanisms. J.Z. and Q.L. performed DFT calculation. All authors analysed the data. H.X. and F.M. wrote the manuscript. F.M. conceived the idea and designed the overall research. F.M. and D.Z. supervised the whole project.

\bibliography{sn-bibliography}

	\section*{Method}
	\subsection{The automation platform for TLC data collection}
	In this work, an automated TLC analysis system (Auto-TLC system) that can automatically complete the entire TLC analysis process is developed. The core of Auto-TLC system is two collaborative robots DOBOT MG400 robot equipped with a capillary and  AUBO i5 robot equipped with a mechanical gripper and a sucker driven by air pump. The solution sample stored on the sample tray is dipped by the capillary and then spotted onto the TLC plates placed on the rack. It is worth noting that the special rack is customized and can place several TLC plates at the same time for high-throughput experiment. Then, the AUBO i5 robot uses the gripper to transfer the spotted TLC plates to a TLC chamber. The sucker is employed to open or close the chamber lid. It is worth noting that there are multiple TLC chambers here, which can be utilized with different elution solvents at the same time, which greatly improves the efficiency of the system. When the developing time reaches the designed value (300 seconds in this work), AUBO i5 robot will retrieve corresponding TLC plate from the chamber and send it to the ultraviolet and visible light camera devices respectively to take photos and record the result. The visible light camera device is utilized to record the position of the frontier of the developing eluent and the ultraviolet camera device is employed to visualize and photograph the spot on the TLC plate. The used TLC plate is dropped into the garbage can and pristine TLC plates are retrieved from storage via the sucker. This cycle will continue until all samples have been tested. 
	
	\subsection{The image recognition algorithm for calculating the $R_f$ values} \label{cal_Rf}
\begin{figure}[h] 
	\centering
	\includegraphics[scale=1.2]{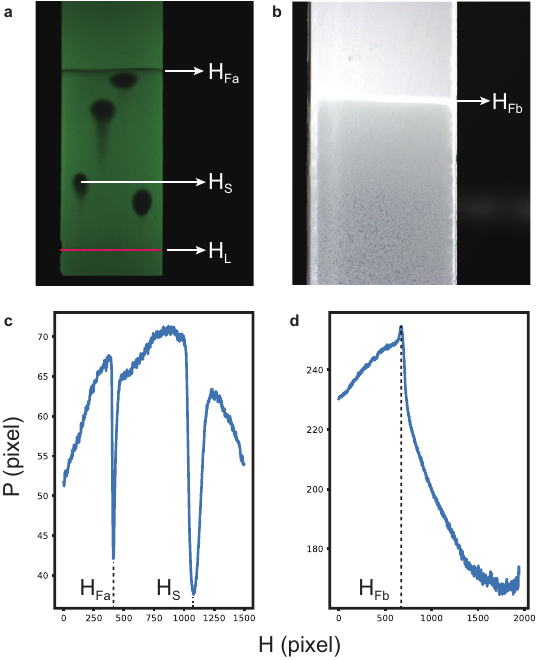}
	\caption{\textbf{Calculation of the $R_f$ values. a,} and \textbf{b,} The photos photographed for the same TLC plate under ultraviolet light and visible light, respectively. \textbf{a} is photographed from the frontal angle and \textbf{b} is photographed from the back angle. \textbf{c,} The average pixel value after y-axis projection performed on the neighborhood of the first TLC spot. \textbf{d,} The average pixel value after y-axis projection performed on the TLC plate photoed under visible light. In this figure, $\mathrm{H}_{\mathrm{Fa}}$ and $\mathrm{H}_{\mathrm{Fb}}$ refers to the height of solvent front in both photos respectively, $\mathrm{H}_{\mathrm{L}}$ refers to the initial height of the TLC spot, $\mathrm{H}_{\mathrm{s}}$ refers to the height of TLC spot and $x_s$ represents the horizontal position of the TLC spot. H is the position of pixel and P is the average pixel value after y-axis projection.}
	\label{figS1} 
	\end{figure}

	In the Auto-TLC system, the $R_f$ values are calculated automatically from an image recognition algorithm based on the recorded photos, which is shown in Figure~\ref{figS1}. For each experiment, four samples are spotted on one TLC plate and two photos are taken from ultraviolet and visible light camera devices, respectively. Benefited from the Auto-TLC analysis platform, the TLC plate is spotted standardized so that the initial position of TLC spot ($x_s$, $\mathrm{H}_{\mathrm{L}}$) ($s=1,2,3,4$) fixed beforehand. In order to calculate the $R_f$ value, the height of each TLC spot $\mathrm{H}_{\mathrm{s}}$, and the height of solvent front $\mathrm{H}_{\mathrm{Fa}}$ and $\mathrm{H}_{\mathrm{Fb}}$ are identified by the projection method. Prior to the projection, the main part of the TLC board can be easily distinguished from the black background by the threshold method. In order to prevent the four TLC spots from interfering with each other during recognition, the projection is made merely in the neighborhood of each TLC spot $[x_s-\Delta x,x_s+\Delta x]$, and the size of the neighborhood $\Delta x$ is selected according to the distance between the points which is fixed in advance. The projection curve of the first TLC spot is taken as an example in Figure~\ref{figS1}c. It is obvious that the curve has two minimum values and $\mathrm{H}_{\mathrm{Fa}}$ and $\mathrm{H}_{\mathrm{s}}$ can be identified from them, since only these two positions can appear black under ultraviolet light in the neighborhood, thereby significantly reducing the average pixel value under the y-axis projection. In the same way, under visible light irradiation, the solvent front is the brightest, so its y-axis projection curve has a maximum value at this position, which corresponds to the value of $\mathrm{H}_{\mathrm{Fb}}$. Considering that the solvent front is not apparent in some cases and cannot be accurately identified under ultraviolet light, the $\mathrm{H}_{\mathrm{Fb}}$ identified from the photo taken under visible light can play an important complementary role.
	
	Since the focal lengths of the cameras are determined, the mapping relationship between the pixel distance on the photo and the actual distance can be determined in advance. Here, the mapping functions for ultraviolet and visible light camera device are referred as $f_a$ and $f_b$, respectively. Therefore, the actual height of the solvent front $\mathrm{H}_{\mathrm{F}}$ is expressed as:
	
$$
\mathrm{H}_{\mathrm{F}}=\left\{\begin{array}{c}
	\frac{1}{2}\left(f_{a}\left(\mathrm{H}_{\mathrm{Fa}}\right)+f_{b}\left(\mathrm{H}_{\mathrm{Fb}}\right)\right), \text { if } \frac{\lvert f_{a}\left(\mathrm{H}_{\mathrm{Fa}}\right)-f_{b}\left(\mathrm{H}_{\mathrm{Fb}}\right)\lvert}{f_{b}\left(\mathrm{H}_{\mathrm{Fb}}\right)}<5 \% \\
	f_{b}\left(\mathrm{H}_{\mathrm{Fb}}\right), \text { other condition }
\end{array}\right.
$$

Here, other condition includes the difference between $\mathrm{H}_{\mathrm{Fa}}$ and $\mathrm{H}_{\mathrm{Fb}}$ is large, or $\mathrm{H}_{\mathrm{Fa}}$ is not identified successfully. With the calculated $\mathrm{H}_{\mathrm{s}}$ and $\mathrm{H}_{\mathrm{F}}$, the $R_f$ value can be finally calculated as:

$$
R_{fs}=\frac{f_{a}\left(\mathrm{H}_{\mathrm{s}}\right)-f_{a}\left(\mathrm{H}_{\mathrm{L}}\right)}{\mathrm{H}_{\mathrm{F}}-f_{a}\left(\mathrm{H}_{\mathrm{L}}\right)}
$$

With the image recognition algorithm, $R_f$ values can be identified quickly and automatically. During the experiments, there may emerge some extreme situations that the algorithm is difficult to judge, such as tailing and fusion of two TLC points. As a consequence, after the high-throughput experiment is completed, the entire result will be manually verified by expert's experience to deal with the extreme situations mentioned above to guarantee the accuracy and reliability of the obtained datasets. A program of visual interface is developed that makes manual verification easy by simply clicking on the correct TLC spot, starting point and solvent front on the image and the image recognition algorithm will correct the mistakes automatically. It is worth mentioning that the extreme situations are rare since the correct $R_f$ value can be obtained through image recognition in most experiments.

	\subsection{Dataset}
	\subsubsection{Description of the dataset}
\begin{figure}[h] 
		\centering
		\includegraphics[scale=0.5]{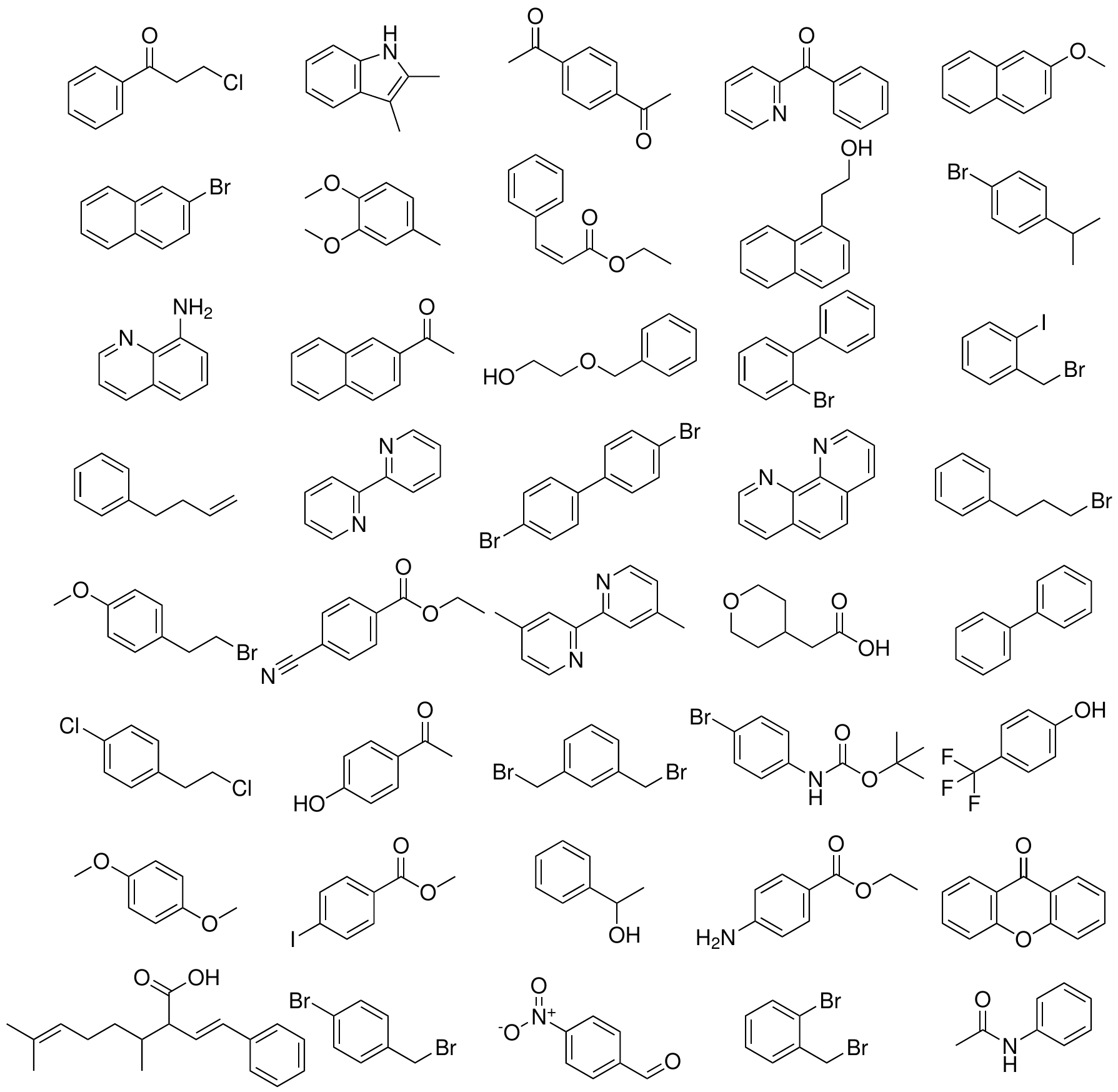}
		\caption{\textbf{Randomly selected compounds in the dataset.}}
		\label{figS2} 
	\end{figure}

	In this work, a valuable dataset of compound polarity is obtained from Auto-TLC analysis platform. In order to ensure the diversity of compounds, the $R_f$ curves of 387 organic compounds, including ketone, aldehyde, ether, halide, alcohol and other categories, under 17 different solvent composition are measured, thereby collecting 4944 standardized polarity measurements. Some of the compounds are shown in Figure~\ref{figS2}. Three elution solvent systems including hexane/ethyl acetate (EA), dichloromethane (DCM)/methanol (MeOH) and hexane/diethyl ether (Et$_2$O) are utilized in this work. For hexane/EA system, eight proportions are employed, namely 1:0, 50:1, 20:1, 5:1, 3:1, 1:1, 1:2 and 0:1. For DCM/MeOH system, six proportions are employed including 1:0, 100:1, 50:1, 30:1, 20:1 and 10:1. For hexane/Et$_2$O system, three proportions are employed including 2:1, 1:1, 0:1.

	\subsubsection{Dataset preprocessing}
	Before machine learning, the dataset acquired from Auto-TLC analysis platform needs to preprocess. Considering that chemical formulas cannot be directly used as input for machine learning, the molecular fingerprint and descriptors are extracted in advance. In this work, the molecular fingerprint Molecular Access System (MACCS) keys are employed to represent molecular structure, fragments or substructure information. MACCSkeys is a one-hot vector with a length of 167 bits, each bit represents a specific molecular structure, 1 represents the structure exists, and 0 represents the structure does not exist. The meaning of each bit can be found in open resource. It is worth mentioning that although each compound corresponds to a unique MACCkeys, each MACCkeys may correspond to several compounds with a similar structure. For example, 134$^{th}$ bit represents the existence of halogen (Cl, Br, I). Therefore, bromobenzene and chlorobenzene has the same MACCkeys since the 134$^{th}$ bit of both is 1 while other substructure is the same. This feature of MACCkeys allows it to extract common features (such as the presence or absence of certain substructure) from countless compounds to predict polarity, thereby achieving the goal of ``small data, big tasks". This is why the polarities of new compounds can be well predicted by only relying on hundreds of different kinds of compounds.
	
	In addition to MACCkeys, several commonly used molecular descriptors including molecular weight (MW), the number of hydrogen bond acceptor (HBA), the number of hydrogen bond donors (HBD), oil-water partition coefficient (LogP), rotatable key (nROTB) and topological polar surface area (TPSA) are employed in this work. These molecular descriptors contain some physical and chemical properties of the molecule, which will contribute to the polarity of the compound. MACCkeys and molecular descriptors utilized in this work can be easily accessed from the python package RDKit, which is a widely utilized tool combining computational chemistry and machine learning. Moreover, another property closely related to polarity, the dipole moment (DM), has also been derived through computational chemistry in this work.
	
	In addition to compound-related information, eluent solvents-related information will also be imported into the machine learning model. Considering that there are multiple developing agent systems in the dataset, a five-length vector [Hexane, EA, DCM, MeOH, Et$_2$O] that represents the proportion of each eluent solvent, is employed to describe the solvent composition. For example, [3, 1, 0, 0, 0] represents the eluent is Hexane/EA = 3:1. With this method, the influence of solvent composition on the $R_f$ values can also be learned through machine learning.
	
	\begin{figure}[h] 
		\centering
		\includegraphics[scale=1]{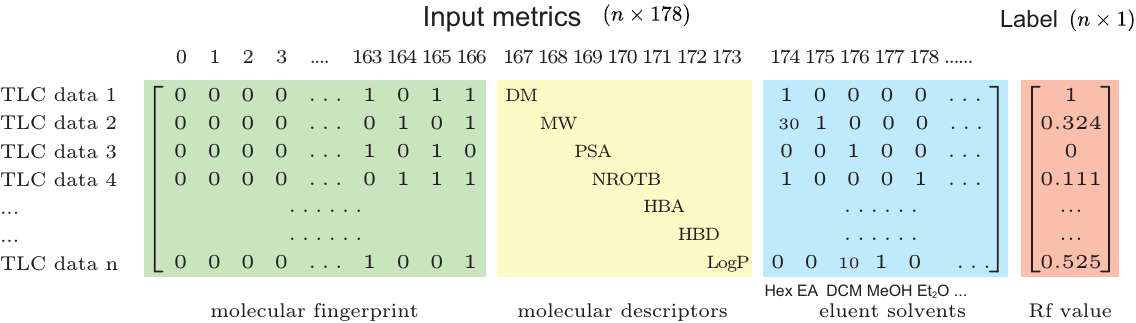}
		\caption{\textbf{The input information matrixs.} Molecular fingerprint, descriptors and elution solvents as input metrics, and $R_f$ value as the label. }
		\label{figS3} 
	\end{figure}
	
	\subsection{Machine learning methods}
	\subsubsection{Random forest (RF)}
	
	Random forest is an algorithm that integrates multiple trees through the idea of ensemble learning. Its basic unit is a decision tree, and its essence belongs to ensemble learning method which is a major branch of machine learning. RF is insensitive to noise in the training dataset, and is more conducive to obtaining a robust model compared with a single decision tree because it uses a set of unrelated decision trees. Meanwhile, it is able to avoid overfitting to a certain extent. In this work, the RF algorithm is implemented by calling the function \textit{RandomForestRegressor} in the python package\textit{ sklearn.ensemble}.
	
	\subsubsection{LightGBM (LGB)}
	
	LGB is constructed on the basis of gradient boosting decision tree (GBDT) algorithm, which improves calculation speed and stability by combining gradient-based one-Side sampling (GOSS) technique and exclusive feature bundling method (EFB). LGB algorithm has satisfactory performance in both regression and classification tasks in the case of large data volume and high-dimensional features. The LGB algorithm has been integrated into the \textit{lightgbm} package and can be employed conveniently in Python.
	
	\subsubsection{Extreme gradient boosting (XGBoost)}
	
	XGBoost is a widely used GBDT-based machine learning method that is able to complete classification and regression tasks efficiently. The optimization goal of XGBoost is to build K regression trees to perform prediction with high accuracy and generalization ability. Therefore, the target loss function can be written as:
	$$
	L=\sum_{i}\left(\hat{y}_{i}-y_{i}\right)^{2}+\sum_{j} \Omega\left(f_{j}\right)
	$$
	where
	$$
	\Omega(f)=\gamma T+\frac{1}{2} \lambda\|w\|^{2}
	$$
	Here,   $\hat{y}_{i}$ is the prediction value, ${y}_{i}$ is the true value; $\Omega(f_{j})$ is the complexity of each regression tree, where $T$ represents the number of leaf nodes and $w$ represents the value of the node. The loss function $L$ can be optimized by greedy algorithm. XGBoost algorithm can be quickly implemented by calling the python package \textit{xgboost}.
	
	\subsubsection{Artificial neural network (ANN)}
	In this work, a feed forward fully-connected ANN is utilized to perform prediction, which comprises an input layer, an output layer, and one or several layer(s) between the input and output layers that are termed hidden layer(s). Each hidden layer is composed of multiple neurons. Two adjacent layers are connected as follows:
	$$
	z_{l}=\sigma\left(W_{l} \vec{z}_{l-1}+\vec{b}_{l}\right), l=1, \ldots, L-1
	$$
	where $l$ denotes the layer index; $W$ denotes the weight matrix; $\vec{b}$ denotes the bias vector; and $\sigma$ denotes the activation function. Consequently, using a neural network approximation, the relationship between the input vector  $\vec{z}_{0}$ and output prediction  $\vec{z}_{L}$ can be expressed as:
	$$
	\vec{z}_{L}=N N\left(z_{0} ; \theta\right)=W_{L} \sigma\left(\cdots \sigma\left(W_{2} \sigma\left(W_{1} \vec{z}_{0}+\vec{b}_{1}\right)+\vec{b}_{2}\right)\right)+\vec{b}_{L}
	$$
	where $\theta$ denotes the collection of all learnable coefficients, which can be written as:
	$$
	\theta=\left\{W_{1}, b_{1}, W_{2}, b_{2}, \ldots, W_{L}, b_{L}\right\}
	$$
	For learning the underlying relationship between the compound properties and polarity, the input vector comprises the information about the structure, properties and eluent solvents, which is represented as $\vec{x}$; and the corresponding $R_f$ value can be termed as $u(\vec{x})$.
	
	Suppose that there are $N$ TLC data, $\left\{u\left(\vec{x}_{i}\right)\right\}_{i=1}^{N}$. In order to train the neural network, a loss function is then defined as follows:
	$$
	\operatorname{Loss}(\theta)=\frac{\sum_{i=1}^{N}\left[u\left(\vec{x}_{i}\right)-N N\left(\vec{x}_{i} ; \theta\right)\right]^{2}}{N}
	$$
	In this work, the Adam optimizer is utilized to minimize the loss function for training the neural network.
	
	\subsubsection{The ensemble method}
	In this work, an ensemble method is proposed to improve the stability and accuracy of the prediction performance. The formula of our proposed ensemble method is written as:
	$$
	u_{\mathrm{pred }}=w_{\mathrm{RF}} u_{\mathrm{RF}}+w_{\mathrm{LGB}} u_{\mathrm{LGB}}+w_{\mathrm{XGB}} u_{\mathrm{XGB}}+w_{\mathrm{ANN}} u_{\mathrm{ANN}}
	$$
	$$
	w_{\mathrm{RF}}+w_{\mathrm{LGB}}+w_{\mathrm{XGB}}+w_{\mathrm{ANN}}=1
	$$
	where $u_{\mathrm{pred }}$ is the prediction of the ensemble method; $ u_{\mathrm{RF}}$, $u_{\mathrm{LGB}}$, $u_{\mathrm{XGB}}$ and $u_{\mathrm{ANN}}$ are the prediction of RF, LGB, XGBoost and ANN, respectively.	$w_{\mathrm{RF}}, w_{\mathrm{LGB}}, w_{\mathrm{XGB}}, w_{\mathrm{ANN}}$ are the respective weights. 
	
	\subsubsection{The hyperparamters utilized in this work}
	There are many hyperparameters used in machine learning methods, which are often selected based on the experience of scientists and trail-and-errors to search for a better model. Therefore, the selection of hyperparameters is of great importance to the repeatability of research. Here, the utilized hyperparamters in this work that are selected by grid-search and repeated trails, are presented below.
	
	For XGBoost, number of estimators is 200, maximum depth is 3 and the learning rate is chosen to be 0.1.
	
	For LGB, number of estimators is 1000, random state is 1 and the number of jobs is chosen to be 1.
	
	For ANN, the number of hidden layers is 4 with 128 neurons in each hidden layer, the number of input neurons is 178 and the number of output neurons is 1. The optimizer is chosen to be Adam with the learning rate 0.005. The maximum training epoch is 5000 and the early stop technique is adopted to prevent overfitting on the basis of the training loss and validating loss.
	
	In addition, in order to avoid the influence of the selection of random seeds, different random seeds are used to conduct multiple experiments and take the statistical average.
	
	\subsection{Experiments}
	\subsubsection{The influence of Sigmoid function}
\begin{figure}[h] 
		\centering
		\includegraphics[scale=0.6]{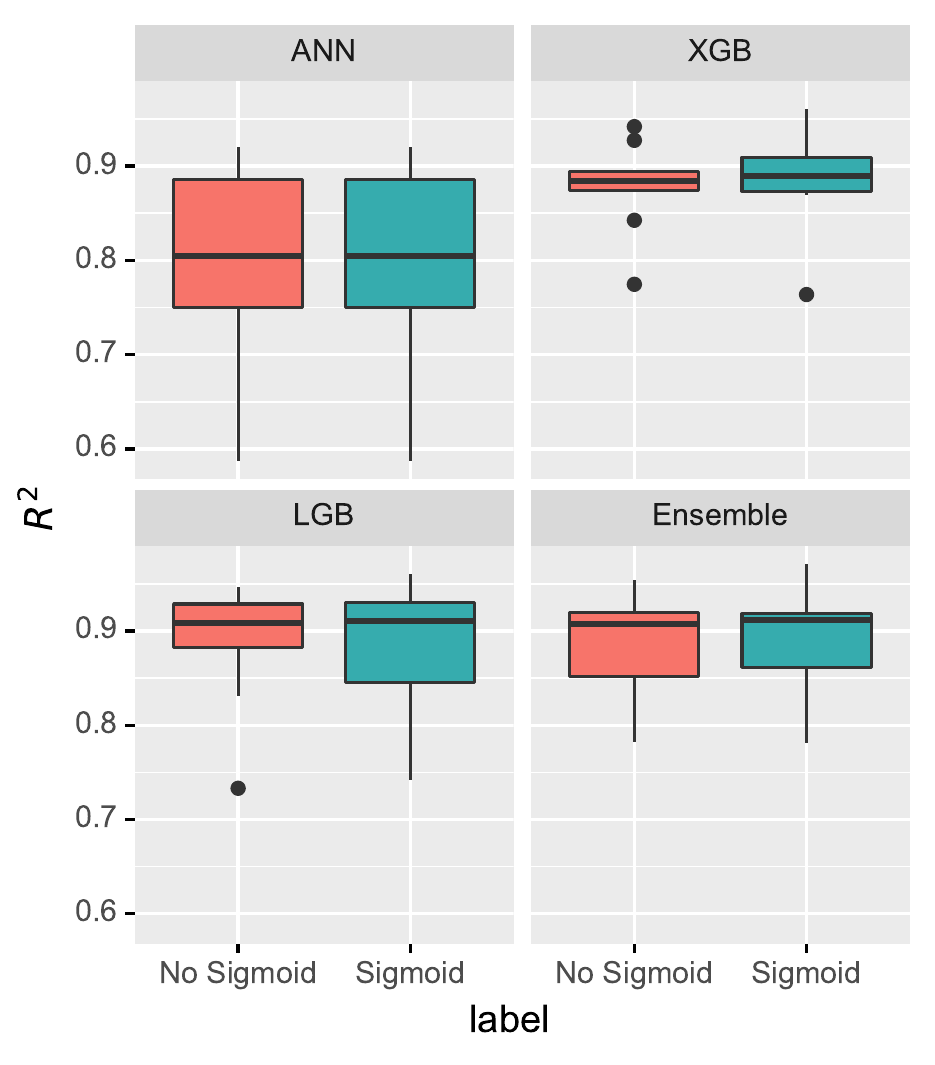}
		\caption{\textbf{The box plot of the performance of different machine learning techniques with or without associating to the Sigmoid function.} The red box plot refers to the result of machine learning methods without employing the Sigmoid function while the green one refers to that employ the Sigmoid function as the physical constraint.}
		\label{figS4} 
	\end{figure}

	Considering the physical constraint that the $R_f$ values is between 0 and 1, which is right with the domain of the Sigmoid function, the formula of which is noted as:
	$$
	f(x)=\frac{1}{1+e^{-x}}
	$$
	As a consequence, in this work, the machine learning techniques are employed to learn a relationship between the input information and the output, which is then mapped to the value domain of (0,1) through the Sigmoid function. In order to better demonstrate the influence of the Sigmoid function, the performance of different machine learning techniques with or without associating to the Sigmoid function are compared, and the result is shown in Figure~\ref{figS4}. It seems that employing the Sigmoid function as the physical constraint is able to improve the stability and accuracy of prediction, especially for the algorithms based on the GBDT like XGB and LGB. Meanwhile, for ANN and the ensemble methods, the performances are also enhanced a little. This indicates that the association of the Sigmoid function is meaningful and effective for different machine learning techniques.

	\subsubsection{The influence of the input information}
	In this work, the input information contains molecular fingerprints that represent the compound structure, molecular descriptors that describe the compound properties,  dipole moment that measures the polarity of a charge system, and the eluent solvents that provide comparable polar environment. In this part, an ablation experiment is taken to investigate the influence of the input information via utilizing different input information to train the machine learning model. The dataset is divided by compound. XGBoost is utilized here as an example. The result is shown in Table~\ref{table1}. Here, merely employing the molecular fingerprint is utilized as the baseline. From the table, it is obvious that adding dipole moment and molecular descriptors is able to improve the prediction performance. It is worth noting that the addition of dipole moment significantly improves the results by 0.0192 compared with the baseline, which means that the dipole moment implies much information about polarity. Therefore, the dipole moment is of great importance for the predicted polarity. However, it can only be obtained from the quantum chemistry calculation that is computationally expensive. Therefore, we provide a valuable dataset with the dipole moment information in this work.
	
	\begin{table}[h]
	\centering
	\caption{\textbf{The $R^2$ of the model trained with XGBoost employing different sections of the input information.} The sign $\checkmark$ represents that corresponding section is contained by the input information and the sign $\times$ represents that corresponding section is not included.}
	\begin{tabular}{ccccc}
		\hline \multicolumn{5}{c}{ XGB } \\
		\hline Fingerprint & $\checkmark$ & $\checkmark$ & $\checkmark$ & $\checkmark$ \\
		Dipole moment & $\checkmark$ & $\times$ & $\checkmark$ & $\times$ \\
		Molecular descriptors & $\checkmark$ & $\checkmark$ & $\times$ & $\times$ \\
		$R^2$ & $0.8811$ & $0.8689$ & $0.8797$ & $0.8605$ \\
		\hline
	\end{tabular}
	\label{table1}
	\end{table}
	
	\subsubsection{Comparison between different machine learning techniques}
	In the paper, the coefficient of determination   is utilized to measure the prediction accuracy. In this section, several other metrics including mean square error (MSE), root mean square error (RMSE) and mean absolute error (MAE) are used to show the differences between different machine learning methods more comprehensively. The dataset is divided by compound. The calculation formulas are written as:
	$$
	\mathrm{MSE}=\frac{\sum_{i}^{N}\left(y_{i}-\hat{y}_{i}\right)^{2}}{N}
	$$
	
	$$
	\mathrm{RMSE}=\sqrt{\frac{\sum_{i}^{N}\left(y_{i}-\hat{y}_{i}\right)^{2}}{N}}
	$$
	
	$$
	\mathrm{MAE}=\frac{\sum_{i}^{N}\lvert y_{i}-\hat{y}_{i}\lvert}{N}
	$$
	
	$$
	R^{2}=\frac{\sum_{i}^{N}\left(y_{i}-\hat{y}_{i}\right)^{2}}{\sum_{i}^{N}\left(y_{i}-\bar{y}_{i}\right)^{2}}
	$$
	where $N$ is the number of test samples, $y_i$ is the true $R_f$ value, $\bar{y}_{i}$ is the mean of true values and $\hat{y}_{i}$ is the predicted $R_f$ value. The performance of machine learning techniques employed in this work is measured by the metrics mentioned above and the result is shown in Table~\ref{table2}. From the table, the difference of different machine learning techniques is obvious and it is discovered that the ensemble method is more accurate and stable.
	
	\begin{table}[h]
	\caption{\textbf{The MSE, RMSE, MAE and $\mathbf{R^2}$ of the prediction made by XGB, LGB, ANN and Ensemble method, respectively.} The results are displayed in the form of mean ± standard deviation, which are calculated from 10 independent trails with different random seeds.}
	\begin{tabular}{ccccc}
		\hline & XGB & LGB & ANN & Ensemble \\
		\hline MSE & $0.0133 \pm 0.0064$ & $0.01366 \pm 0.0077$ & $0.0233 \pm 0.0105$ & $\mathbf{0 . 0 1 2 4} \pm \mathbf{0 . 0 0 6 5}$ \\
		RMSE & $0.1123 \pm 0.0257$ & $0.1128 \pm 0.0307$ & $0.1486 \pm 0.0350$ & $\mathbf{0 . 1 0 7 7} \pm \mathbf{0 . 0 2 8 1}$ \\
		MAE & $0.0756 \pm 0.0174$ & $0.0748 \pm 0.0206$ & $0.0990 \pm 0.0220$ & $\mathbf{0 . 0 7 3 5} \pm \mathbf{0 . 0 1 9 7}$ \\
		$R^2$ & $0.8864 \pm 0.0493$ & $0.8817 \pm 0.0648$ & $0.7955 \pm 0.1025$ & $\mathbf{0 . 8 9 2 5} \pm \mathbf{0 . 0 5 4 5}$ \\
		\hline
	\end{tabular}
	\label{table2}
	\end{table}

	\subsubsection{Performance of predicted model trained with sparse data}
	In this part, different numbers of compounds including 9(2.5\%), 19(5\%), 38(10\%), 96(25\%), 116(30\%), 193(50\%), 270(70\%) are randomly selected to form the training dataset and the performance of the ensemble prediction model trained from respective training data is examined. The result is shown in Figure~\ref{fig3}e. It is surprised to discover that the accuracy of the prediction model increases rapidly until the training data reaches 30\% of the entire compounds (116 compounds) and keeps high with more training data. It means that less data is required to learn the constitutive relationship between structure and properties of the compound and $R_f$ values, which shows the powerful learning ability of the machine learning techniques.
	
	\subsubsection{Application in selecting elution solvents}
\begin{figure}[h] 
		\centering
		\includegraphics[scale=0.5]{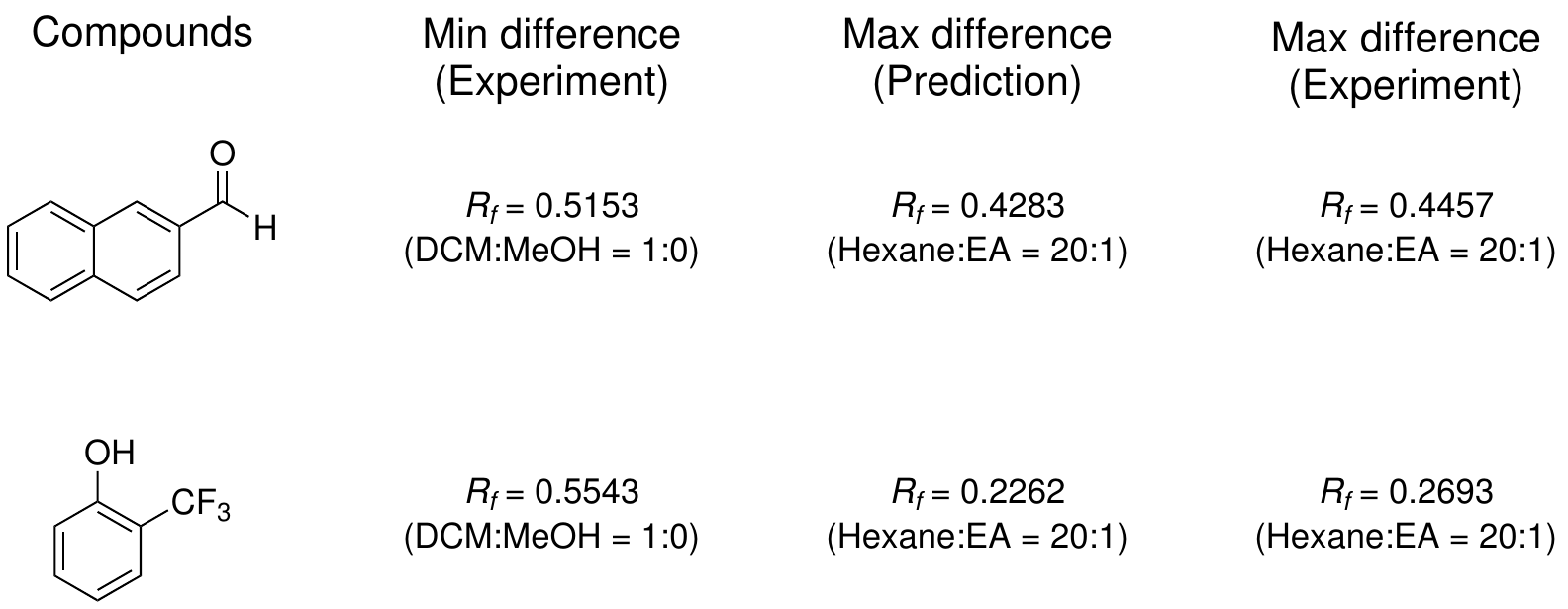}
		\caption{\textbf{An example of selecting elution solvents from prediction model.} Min difference (Experiment) and Max difference (Experiment) refer to the elution solvents and corresponding $R_f$ values with the smallest and largest $R_f$ values difference obtained from multiple experiments, respectively. Max difference (prediction) represents the elution solvents and corresponding $R_f$ values with the largest $R_f$ values difference through the prediction model.}
		\label{figS5} 
	\end{figure}

	In TLC analysis, selecting a suitable elution solvents is the most difficult task and relies heavily on researchers' experience. In some cases, the two compounds may have little difference in $R_f$ values under the same solvent system, resulting inseparability. Therefore, it is necessary to find a suitable solvent system to maximize the difference between the $R_f$ values of the two compounds at this time. Under normal circumstances, this process requires repeated experimentation of different solvent systems, which is time-consuming and labor-intensive. With the predictive model, it is easy to find the optimal elution solvents by comparing the predicted values under different these conditions, which greatly improves the work efficiency. An example is provided in Figure~\ref{figS5}. It is found that the difference between the two compound is very small under DCM:MeOH system, while the difference is obvious in Hexane:EA system. Meanwhile, the $R_f$ value under the optimal elution solvents found by prediction model is parallel with the result of repeated experiments, which shows the accuracy and power of prediction model.
	
	\subsection{Mathematical derivation of TLC adsorption model}
\begin{figure}[h] 
		\centering
		\includegraphics[scale=\aaa]{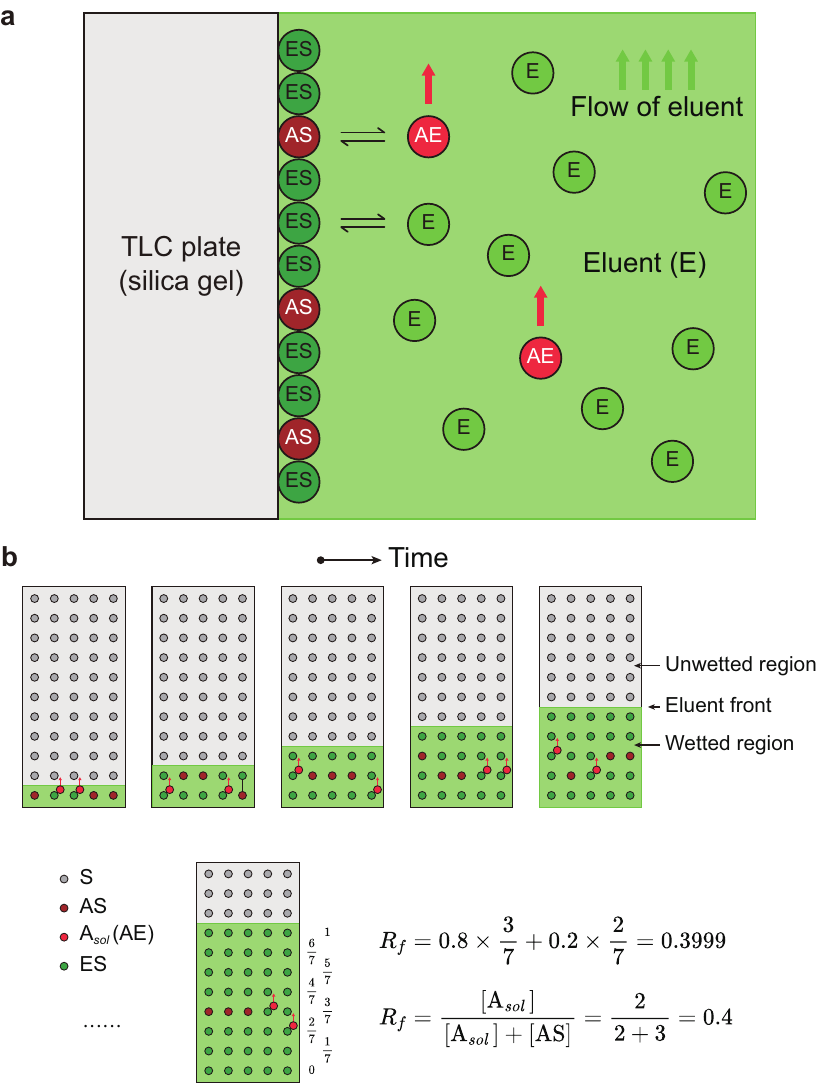}
		\caption{\textbf{Adsorption model of a TLC plate. a,} Adsorption equilibrium of solute (\textbf{A}) and eluent (\textbf{E}) on the surface of silica gel. \textbf{b,} Relation of $R_f$ value and adsorption equilibrium.}
		\label{figS6} 
	\end{figure}

	We put forward a mathematic model of adsorption equilibria on the surface of silica gel. As the eluent flows on the TLC plate, both the solute molecule \textbf{A} and eluent molecule \textbf{E} generate a fast adsorption-desorption equilibrium on the surface of silica gel (Figure~\ref{figS6}a). The adsorbed solute molecule, \textbf{AS}, will keep stationary on the TLC plate, while the desorbed solute molecule dissolved in the eluent, \textbf{AE}, will move together with the eluent.
	
	Accordingly, a TLC process was imitated as shown in Figure~\ref{figS6}b. For the sake of presentation, we built a 5 $\times$ 11 matrix of adsorption sites of silica gel on the TLC plate. At the beginning, 5 solute molecules \textbf{A} stand on the starting line. In this case, we suppose that the adsorption equilibrium is [\textbf{AE}]/[\textbf{AS}] = 2/3. At each time point, 2 solute molecules were picked randomly as \textbf{AE} and the other 3 as \textbf{AS}. Due to the fast equilibration of adsorption and desorption, the solute molecules are rapidly transformed between \textbf{AE} and \textbf{AS}. The solvent front moved upward for one frame and the desorbed \textbf{AE} moved together, while the adsorbed \textbf{AS} remained stationary. The ratio in height between the center of the \textbf{AS} distributed region and the solvent front, i.e. the $R_f$ value, will be the ratio of solvated \textbf{A}:
	$$
	R_{f}=\frac{[\mathrm{AE}]}{[\mathrm{AE}]+[\mathrm{AS}]}
	$$
	
	$$
	\mathrm{A}_{i}+\mathrm{E}_{j} \rightleftharpoons \mathrm{A}_{i} \mathrm{E}_{j} 
	$$
	
	$$
	K_{\mathrm{A E}, i j}=\frac{\left[\mathrm{A}_{i} \mathrm{E}_{j}\right]}{\left[\mathrm{A}_{i}\right]\left[\mathrm{E}_{j}\right]}
	$$
	
	$$
	\mathrm{A}_{i}+\mathrm{S} \rightleftharpoons \mathrm{A}_{i} \mathrm{S}
	$$
	
	$$
	K_{\mathrm{A S}, i}=\frac{\left[\mathrm{A}_{i} \mathrm{S}\right]}{\left[\mathrm{A}_{i}\right][\mathrm{S}]}
	$$
	
	$$
	\mathrm{E}_{j}+\mathrm{S} \rightleftharpoons \mathrm{E}_{j} \mathrm{S}
	$$
	
	$$
	K_{\mathrm{E S}, j}=\frac{\left[\mathrm{E}_{j} \mathrm{S}\right]}{\left[\mathrm{E}_{j}\right][\mathrm{S}]}
	$$
	
	$$
	\left[\mathrm{A}_{i, \mathrm{sol}}\right]=\left[\mathrm{A}_{i}\right]+\sum_{j=1}^{n}\left[\mathrm{A}_{i} \mathrm{E}_{j}\right]
	$$
	
	$$
	\left[\mathrm{A}_{i} \mathrm{E}_{j}\right]=K_{\mathrm{A E}, i j}\left[\mathrm{E}_{j}\right]\left[\mathrm{A}_{i}\right]
	$$
	
	$$
	\left[\mathrm{A}_{i, \text {sol }}\right]=\left[\mathrm{A}_{i}\right]\left(1+\sum_{j=1}^{n} K_{\mathrm{AE}, i j}\left[\mathrm{E}_{j}\right]\right)
	$$
	
	$$
	\mathrm{S}_{0}=[\mathrm{S}]+\sum_{i=1}^{m}\left[\mathrm{A}_{i} \mathrm{S}\right]+\sum_{j=1}^{n}\left[\mathrm{E}_{j} \mathrm{S}\right]
	$$
	
	$$
	\left[\mathrm{A}_{i} \mathrm{S}\right]=K_{\mathrm{AS}, i}\left[\mathrm{A}_{i}\right][\mathrm{S}]
	$$
	
	$$
	\left[\mathrm{E}_{j} \mathrm{S}\right]=K_{\mathrm{ES}, j}\left[\mathrm{E}_{j}\right][\mathrm{S}]
	$$
	
	$$
	\mathrm{S}_{0}=[\mathrm{S}]\left(1+\sum_{i=1}^{m} K_{\mathrm{AS}, i}\left[\mathrm{A}_{i}\right]+\sum_{j=1}^{n} K_{\mathrm{ES}, j}\left[\mathrm{E}_{j}\right]\right)
	$$
	
	$$
	R_{f,i}=\frac{\left[\mathrm{A}_{i, sol}\right]}{\left[\mathrm{A}_{i, sol}\right]+\left[\mathrm{A}_{i} \mathrm{S}\right]}=\frac{1}{1+\left[\mathrm{A}_{i} \mathrm{S}\right] /\left[\mathrm{A}_{i, sol}\right]}
	$$
	
	$$
	\frac{\left[\mathrm{A}_{i} \mathrm{S}\right]}{\left[\mathrm{A}_{i, sol}\right]}=\frac{\left[\mathrm{A}_{i} \mathrm{S}\right]}{\left[\mathrm{A}_{i}\right]} \frac{\left[\mathrm{A}_{i}\right]}{\left[\mathrm{A}_{i, sol}\right]}=K_{\mathrm{AS}, i}[\mathrm{S}] \frac{1}{1+\sum_{j=1}^{n} K_{\mathrm{AE},ij}\left[\mathrm{E}_{j}\right]}
	$$
	$$
	=\frac{K_{\mathrm{AS}, i} \mathrm{S}_{0}}{\left(1+\sum_{i=1}^{m} K_{\mathrm{AS}, i}\left[\mathrm{A}_{i}\right]+\sum_{j=1}^{n} K_{\mathrm{ES}, j}\left[\mathrm{E}_{j}\right]\right)\left(1+\sum_{j=1}^{n} K_{\mathrm{AE}, i j}\left[\mathrm{E}_{j}\right]\right)}
	$$
	
	$$
	R_{f,i}=\frac{1}{1+K_{\mathrm{A S}, i} \mathrm{S}_{0} /\left(1+\sum_{i=1}^{m} K_{\mathrm{A S}, i}\left[\mathrm{A}_{i}\right]+\sum_{j=1}^{n} K_{\mathrm{E S}, j}\left[\mathrm{E}_{j}\right]\right)\left(1+\sum_{j=1}^{n} K_{\mathrm{A E}, i j}\left[\mathrm{E}_{j}\right]\right)}
	$$
	
	$$
	1+\sum_{i=1}^{m} K_{\mathrm{A S}, i}\left[\mathrm{A}_{i}\right]+\sum_{j=1}^{n} K_{\mathrm{E S}, j}\left[\mathrm{E}_{j}\right] \approx \sum_{j=1}^{n} K_{\mathrm{E S}, j}\left[\mathrm{E}_{j}\right]
	$$
	
	$$
	1+\sum_{j=1}^{n} K_{\mathrm{A E}, i j}\left[\mathrm{E}_{j}\right] \approx \sum_{j=1}^{n} K_{\mathrm{A E}, i j}\left[\mathrm{E}_{j}\right]
	$$
	
	$$
	R_{f,i}=\frac{1}{1+K_{\mathrm{A S}, i} \mathrm{S}_{0} /\left(\sum_{j=1}^{n} K_{\mathrm{E S}, j}\left[\mathrm{E}_{j}\right]\right)\left(\sum_{j=1}^{n} K_{\mathrm{A E}, i j}\left[\mathrm{E}_{j}\right]\right)}
	$$
	Through mathematical model derivation, we can prove that the $R_f$ value is determined by the following equation, where $\mathrm{S}_{0}$ is the total amount of adsorption sites of silica gel.
	$$
	R_f=\frac{1}{1+\left(\frac{K_{\mathrm{A S}}}{K_{\mathrm{E S}} K_{\mathrm{A E}}}\right)\left(\frac{\mathrm{S}_{0}}{[\mathrm{E}]^{2}}\right)}
	$$
	By comparing it with Sigmoid function:
	$$
	f(x)=\frac{1}{1+e^{-x}}
	$$
	We find that the independent variable of Sigmoid, $x$, represents the change of energy between the competitive adsorption of \textbf{A} and \textbf{E}:
	$$
	x=-\ln \left(\frac{K_\mathrm{A S}}{K_\mathrm{E S} K_\mathrm{A E}} \frac{\mathrm{S}_{0}}{[\mathrm{E}]^{2}}\right)=c+\frac{\Delta G_{\mathrm{A}, ads}^{\Theta}-\Delta G_{\mathrm{E}, ads}^{\Theta}-\Delta G_{\mathrm{A E}, sol}^{\Theta}}{R T}
	$$
	In other words, the $R_f$ value is a Sigmoid function of the change of Gibbs free energy for the competitive adsorption reaction:
	$$
	\mathrm{AE}+\mathrm{ES} \rightleftharpoons \mathrm{AS}+2 \mathrm{E}
	$$
	
	$$
	\Delta G_\mathrm{A S-E S}^{\Theta}=\Delta G_{\mathrm{A}, ads}^{\Theta}-\Delta G_{\mathrm{E}, ads}^{\Theta}-\Delta G_{\mathrm{A E}, sol}^{\Theta}
	$$
	
	$$
	R_f=\frac{1}{1+e^{-\left(c+\frac{\Delta G_\mathrm{A S-E S}^{\Theta}}{R T}\right)}}
	$$
	Where the constant $c$ is based on the concentration difference between the eluent and the adsorption sites on the silica gel.

\end{document}